\documentclass[11pt]{article}
\newcommand{\bm}[1]{\mbox{\boldmath $#1$}}
\renewcommand{\theequation}{\arabic{section}.\arabic{equation}}
 \def\bb{\bibitem} \def\lb{\label}
\def\be{\begin{equation}} \def\ee{\end{equation}}
\def\ba{\begin{eqnarray}} \def\ea{\end{eqnarray}} \def\part{\partial}
\def\nn{\nonumber}
\def\ol{\overline} \def\z{\zeta}
\def\L{\Lambda} \def\R{{\cal R}}
\def\X{{\bm X}} \def\J{{\bm J}}
\def\a{{\bm \alpha}} \def\b{{\bm \beta}} \def\c{{\bm \gamma}}

\begin{document}

\begin{titlepage}

\title{\begin{flushright}\begin{small}    LAPTH-1326/09
\end{small} \end{flushright} \vspace{1.5cm}
Black holes with a null Killing vector in\\ three-dimensional
massive gravity}
\author{G\'erard Cl\'ement \thanks{Email:
gclement@lapp.in2p3.fr}\\ \\
{\small Laboratoire de  Physique Th\'eorique LAPTH (CNRS),} \\
{\small B.P.110, F-74941 Annecy-le-Vieux cedex, France}}

\date{5 May 2009}
\maketitle

\begin{abstract}
We investigate solutions of new massive gravity with two commuting
Killing vectors, one of which is null, with a special emphasis on
black hole solutions. Besides extreme BTZ black holes and, for a
special value of the coupling constant, massless null warped black
holes, we also obtain for a critical coupling a family of massive
``log" black holes. These are asymptotic to the extreme BTZ black
holes in the sense of log gravity.
\end{abstract}

\end{titlepage}\setcounter{page}{2}

\section{Introduction}
It is well known that three-dimensional cosmological Einstein
gravity is dynamically trivial, without propagating degrees of
freedom. This is cured in topologically massive gravity (TMG)
\cite{djt} by the addition to the Einstein action of a
parity-violating gravitational Chern-Simons term, leading after
linearization to the propagation of massive quanta. The same goal is
achieved in a parity-preserving fashion in the recently proposed new
theory of massive gravity (NMG) \cite{bht} through the addition of a
particular quadratic combination of the curvature tensor components.
This theory was found to be unitary in the tree-level in \cite{no},
and renormalizable in \cite{renorm}. New massive gravity was shown
in \cite{newmass} to admit both BTZ black holes \cite{btz} and
warped $AdS_3$ black holes \cite{tmgbh,tmgebh,ALPSS} as solutions,
and the entropy, mass and angular momentum of these black holes were
computed. The central charges for these black holes were recently
obtained in \cite{ks}.

In cosmological TMG linearized around a constant curvature
background, either the massive gravitons or the BTZ black holes have
negative energy, except for a critical, ``chiral'' value of the
Chern-Simons coupling constant at which all the masses vanish
\cite{chiral}. However it was shown in \cite{gaston} that for this
special value TMG also admit a family of massive black holes with
the extremal BTZ black hole as ground state. These exact ``log"
solutions are a limiting case of a family of exact solutions of
cosmological TMG first constructed in \cite{part}.

Similarly, the sign of the energy of massive excitations of NMG
linearized around an $AdS_3$ background is opposite to the sign of
the mass of the BTZ black holes, except for the critical value $m^2
= 1/2l^2$ of the quadratic coupling constant, at which all the
masses again apparently vanish \cite{liusun, liusun2}
\footnote{Actually it has recently been shown \cite{bht2} that the 
situation is more subtle, the massless gravitons being replaced for 
$m^2 = 1/2l^2$ by massive photons.}. The purpose of this paper is to
investigate whether the special massive black hole solutions of
\cite{gaston} can be generalized to critical NMG.

The solutions of \cite{gaston} are actually a special case of $AdS$
wave solutions of TMG \cite{ah,gps,cdww}, reinterpreted as black
hole solutions. As the present work was under way, the paper
\cite{gaston2} came out, in which $AdS$ wave solutions of NMG are
constructed and studied. Our solutions are also special cases of the
solutions of \cite{gaston2}, however the interpretation is
different.

In the next section we construct, using the methods of \cite{part}
and \cite{newmass}, solutions of NMG with two commuting Killing
vectors, one of which is null. These solutions exist for all real
values of the coupling contant $m^2$, however, as shown in Sect. 3
based on the results of the Appendix, they lead to regular black
holes (other than the extreme BTZ black holes) only for the three
values $m^2l^2 = 17/2$, $m^2l^2 = 7/2$ and $m^2l^2 = 1/2$. The
entropy, mass and angular momentum of these black holes are computed
in Sect. 4. Only the $m^2l^2 = 1/2$ analogs of the black holes of
\cite{gaston} are massive. Our results are briefly discussed in the
last section.

\section{Stationary solutions with a null Killing vector}
The action of the cosmological new massive gravity theory is
\cite{bht} \be\lb{act} I_3 = \frac1{16\pi G}\int d^3x
\sqrt{|g|}\left[\R - \frac1{m^2}K - 2\L \right] \,, \ee where $\R$
is the trace of the Ricci tensor $\R_{\mu\nu}$, the quadratic
curvature invariant $K$ is \be K = \R_{\mu\nu}\R^{\mu\nu} -
\frac38\R^2\,, \ee and $G$, $m^2$ and $\L$ are the Newton constant,
quadratic coupling constant and ``bare" cosmological constant.

Let us briefly recall the dimensional reduction of this theory as
carried out in \cite{newmass} in the case of two commuting Killing
vectors. We choose the parametrisation \be \lb{par}
ds^2=\lambda_{ab}(\rho)\,dx^a dx^b + \zeta^{-2}(\rho)R^{-2}(\rho)
\,d\rho^2\,, \ee ($x^0 = t$, $x^1 = \varphi$), where $\lambda$ is
the $2 \times 2$ matrix \be \lambda = \left(
\begin{array}{cc}
T+X & Y \\
Y & T-X
\end{array}
\right), \ee $R^2 \equiv \X^2$ is the Minkowski pseudo-norm of the
``vector'' $\X(\rho) = (T,\,X,\,Y)$, \be \X^2 = \eta_{ij}\,X^iX^j =
-T^2+X^2+Y^2 \,, \ee and the scale factor $\zeta(\rho)$ allows for
arbitrary reparametrizations of the radial coordinate $\rho$. The
scalar product of two vectors $\X$ and $\bm Y$ is defined by
$\X\cdot{\bm Y} = \eta_{ij}\,X^iX^j$, and their wedge product by \be
({\X} \wedge {\bm Y})^i = \eta^{ij}\epsilon_{jkl}X^k Y^l \ee (with
$\epsilon_{012} = +1$). The ansatz (\ref{par}) reduces the equations
of NMG to \ba\lb{eqX} && \X\wedge(\X\wedge\X'''') +
\frac52\X\wedge(\X'\wedge\X''')
+ \frac32\X'\wedge(\X\wedge\X''') \nn\\
&& \quad + \frac94\X'\wedge(\X'\wedge\X'') - \frac12\X''\wedge(\X\wedge\X'') \\
&& \quad - \left[\frac18(\X'^2) + \frac{m^2}{\z^2}\right]\X'' = 0
\,, \nn \ea
and
\ba\lb{scal}
&& \frac12(\X\cdot\X'')^2 - \frac12(\X^2)(\X''^2) - \frac14(\X'^2)(\X\cdot\X'')
- \frac3{32}(\X'^2)^2 \nn\\
&& - \frac{m^2}{\z^2}\left[\frac32(\X'^2) + 2(\X\cdot\X'')\right] -
\frac{6m^2\L}{\z^4} = 0\,. \ea

The equations (\ref{eqX}) are trivially solved by $\X'' = 0$,
leading to constant curvature spacetimes \be\lb{btz} \X = \b\rho +
\c\,, \ee where $\b$ and $\c$ are two linearly independent constant
vectors, and the scale $\b^2 = b^2$ of $\b$ is related through
(\ref{scal}) to the bare cosmological constant. A strategy to
generate non-trivial solutions of the system
(\ref{eqX})-(\ref{scal}) is to consider linear deformations of the
trivial ansatz (\ref{btz}) which solve these equations exactly.
Warped $AdS_3$ black hole solutions were obtained in \cite{newmass}
from the quadratic ansatz $\X = \a\rho^2 + \b\rho + \c$, with
\be\lb{b} \a\wedge\b = b\a\,, \ee for some real constant $b$,
implying \be \a^2 = 0\,, \quad (\a\cdot\b) = 0\,, \quad \b^2 =
b^2\,. \ee In this paper, we consider the ansatz \cite{part}
\be\lb{sd} \X = \a F(\rho) + \b\rho\,, \ee where the form of the
function $F(\rho)$ shall be determined from the field equations.

Before applying this ansatz to NMG, it is instructive to recall its outcome in
TMG. The field equations of TMG reduced according to the stationary circularly
symmetric ansatz (\ref{par}) are \cite{part}
\ba
\X'' & = & \frac{\z}{2\mu}[3\X'\wedge\X'' + 2\X\wedge\X''']\,, \lb{Xtmg} \\
\X'^2 &+& \frac43(\X\cdot\X'') + \frac{4\L}{\z^2} = 0 \,.
\lb{scaltmg} \ea Provided the vectors $\a$ and $\b$ satisfy
(\ref{b}), the vector equation (\ref{Xtmg}) is linearized by
(\ref{sd}) to \be\lb{mastmg} \rho F''' +\left(\frac32+\frac{\mu}{\z
b}\right)F'' = 0\,, \ee while the scalar equation (\ref{scaltmg})
leads to \be b^2 = \frac4{\z^2 l^2} \ee for a negative cosmological
constant $\L = -l^{-2}$. Without loss of generality we can choose
$\z = 1$ and $b = 2/l$, leading to the solution \cite{part} of the
master equation (\ref{mastmg}) \be\lb{Ftmg} F(\rho) = a\rho^p +
c\rho +d\,, \qquad p = \frac{1-\mu l}2\,, \ee depending on three
integration constants $a$, $c$ and $d$. Actually the constant $c$ is
redundant and may be set to zero by a redefinition of the vector $\b$, 
$\b \to \tilde\b = \b + c\a$, which does not affect Eq. (\ref{b}). For 
$p = 1$ ($\mu l = -1$) or $p = 0$ ($\mu l = +1$) the solution 
(\ref{Ftmg}) with $c=0$ degenerates to \cite{gaston}
\ba F(\rho) &=& a\rho\ln|\rho/\rho_0| + d \,, \quad (\mu l =
-1) \lb{mul-1}\\
F(\rho) &=&  d\ln|\rho/\rho_0|\,, \quad (\mu l =
+1)\lb{chir} \ea

In the case of NMG, the vector equation (\ref{eqX}) is again
linearized by the ansatz (\ref{sd}) (with $a$ and $b$ satisfying
(\ref{b})) to the fourth order equation \be\lb{mast} \rho^2F'''' +
4\rho F''' + \left(\frac{17}8 - \frac{m^2}{\z^2b^2}\right)F'' = 0\,,
\ee the scalar equation (\ref{scal}) leading to the constraint
\be\lb{solscal} \frac{b^4}{32} + \frac{m^2b^2}{2\z^2} +
\frac{2m^2\L}{\z^4} = 0\,. \ee Assuming $b^2 > 0$, we again choose
$\z = 1$ and \be b = \frac2{l}\,, \ee where the effective $AdS_3$
curvature parameter $l$ is  obtained by solving (\ref{solscal}),
\be\lb{ell} l^{-2} = 2m^2\left[-1 \pm \sqrt{1-\L/m^2}\right]\,. \ee
The solution of the master equation (\ref{mast}) then leads to
\be
\lb{Fr} F(\rho) = a_+\rho^{p_+} + a_-\rho^{p_-} + d\,,
\quad p_{\pm} = \frac{1\pm\sqrt{m^2l^2+1/2}}2\,, \ee now depending
on three integration constants $a_+$, $a_-$ and $d$ (again we have 
discarded a redundant term $c\rho$ by a redefinition of the vector $\b$). 
Note that
from (\ref{ell}), \be m^2l^2+\frac12 = \pm m^2l^2\sqrt{1-\L/m^2}\,,
\ee so that the square root in (\ref{Fr}) is real either for the
upper sign in (\ref{ell}) and $m^2 > 0$, $\L < 0$, or for the lower
sign in (\ref{ell}) and $m^2 < 0$, $\L > m^2$. For $m^2/l^2 < -
1/2$, (\ref{Fr}) is replaced by \be\lb{Fi} F(\rho) =
a\rho^{1/2}\sin[\gamma\ln|\rho/\rho_0|] + d\,, \quad \gamma =
\frac12\sqrt{-m^2l^2-1/2} \ee (with $\rho_0$ another integration
constant). For the special value $m^2l^2 = -1/2$ ($\L = m^2$),
(\ref{Fr}) or (\ref{Fi}) degenerate to \be\lb{F0} F(\rho) =
a\rho^{1/2}\ln|\rho/\rho_0| + d\,. \ee Finally, (\ref{Fr}) is
also degenerate for the value $m^2l^2 = +1/2$ ($\L = -3m^2$), where
it must be replaced by \be\lb{F1} F(\rho) = a\rho\ln|\rho/\rho_0| +
d\ln|\rho/\rho_0|\,, \ee or, if $d=0$, \be\lb{F12} F(\rho) =
a\rho\ln|\rho/\rho_0| + \ol{d}\ee ($\ol{d}$ constant).

The choice of basis vectors \be\lb{ab} \a =
\frac12(1+l^2,\,1-l^2,\,-2l)\,, \quad \b =
(1-l^{-2},\,-1-l^{-2},\,0)\,, \ee leads to the metric \be\lb{sd1}
ds^2 = [-2l^{-2}\rho + F(\rho)]\,dt^2 - 2lF(\rho)\,dt\,d\varphi +
[2\rho + l^2F(\rho)]\,d\varphi^2  + \frac{l^2\,d\rho^2}{4\rho^2} \ee
which, in the limiting case of a constant $F(\rho) =$ M/2, reduces
to the extreme BTZ metric with J/$l =$ M. The metric (\ref{sd1}) can
be put in the form \be\lb{sd2} ds^2 = l^2[dx^2 + 2e^{2x}dudv +
h(x)du^2]\,, \ee with $x = (1/2)\ln(\rho/l^2)$, $u = \varphi -
l^{-1}t$, $v = \varphi + l^{-1}t$, $h(x) = F(\rho)$. This has the
obvious Killing vectors $L_1 = \partial_u$ and $L_2 = \partial_v$,
the latter being null. In the generic case these are the only
infinitesimal isometries of (\ref{sd2}). In the case of the solution
(\ref{Fr}) with $a_{\mp} = d = 0$, the metric has a third local
isometry generated by \be L_3 = u\partial_u +
\left(\frac2{p_{\pm}}-1\right)v\partial_v - \partial_{x}\,.
\ee Similarly, in the case of the solution (\ref{F1}) ($p_+=1$) with
$d =0$, the third local Killing vector is \be L_3 =
\partial_u + (al^2u + v)\partial_v - \partial_{x}\,. \ee Finally,
in the special cases $a_+ = a_- = 0$ or $a_- = d = 0$ with $p_+ =
2$, the metric (\ref{sd2}) describes respectively extreme BTZ black
holes or null warped black holes (see next section), which both
admit four local Killing vectors generating the $sl(2,R)\times R$
algebra (the Killing vectors for the null warped black hole case are
given in Eqs. (6.3) and (6.4) of \cite{tmgebh}).

All the results of this section are consistent with the results of
\cite{gaston2}, so that our solutions (\ref{sd2}) with $h(x) =
F(\rho)$ given by (\ref{Fr}) and (\ref{Fi})-(\ref{F12}) are special
cases of the $AdS$ waves \be\lb{pp} ds^2 = l^2[dx^2 + 2e^{2x}dudv +
h(x,u)du^2]\,, \ee of \cite{gaston2}. For these solutions to
describe black holes, the spacelike Killing vector
$\partial_{\varphi}$ should have closed orbits, which essentially
restricts (\ref{pp}) to (\ref{sd2}).

\setcounter{equation}{0}
\section{Black holes}
In the present paper we are interested in regular black hole
solutions. The metric (\ref{sd1}) has a horizon at $\rho = 0$. As
pointed out in \cite{part}, all the scalar curvature invariants
constructed from this metric are constant (the function $F(\rho)$
does not contribute because $\a$ is null and orthogonal to $\b$),
however the metric may develop non-scalar curvature singularities at
the horizon $\rho = 0$, as well as at $\rho = \infty$, or the
horizon can be at geodesic infinity. To elucidate this question, we
consider the first integral of the geodesic equation \be \lb{geo}
\dot\rho^2 - l^2P_+^2F(\rho) - 2P_+P_-\rho +
\frac{4\varepsilon}{l^2}\rho^2 = 0\,, \ee with $P_{\pm} = E \pm
l^{-1}L$, where $E$ and $L$ are the constant conjugate momenta to
$\dot{t}$ and $\dot\varphi$, and $\varepsilon = +1, 0$ or $-1$ for
timelike, null, or spacelike geodesics. In the discussion of this
equation, we can exclude outright the case (\ref{Fi}), for which the
areal radius \be r^2 = 2\rho + l^2F(\rho) \ee oscillates wildly
around zero for $\rho \to 0$, leading to naked closed timelike
curves (CTC). For $a_+=a_-=0$ the metric (\ref{sd1}) reduces under
the radial coordinate transformation $\rho = (r^2-dl^2)/2$ to the 
extreme BTZ black
hole metric with mass parameter M $= 2d$. If $a_+$ and $a_-$ do not
both vanish, a lengthy analysis, carried out in the Appendix, leads
to the conclusion that the metric (\ref{sd1}) leads to regular black
holes in only three cases:

1) For $m^2l^2 = 17/2$, the solution (\ref{Fr}) with $a_- = 0$, $a_+
> 0$, $p_+ = 2$ and $d \ge 0$ yields a regular black
hole (free from naked CTC). After transforming to a rotating frame
in which the null Killing vector is $\partial_t$, which amounts to
replacing the basis vectors (\ref{ab}) by \be\lb{ab1} \a =
\frac{l^2}2(1,\,-1,\,0)\,, \quad \b = (1,\,-1,\,2l^{-1})\,, \ee this
leads to a null warped black hole \cite{tmgebh,ALPSS} (corresponding
to the warped black holes of \cite{newmass} with $\beta^2 = 1$)
\be\lb{null} ds^2 = -\frac{4\rho^2}{l^2r^2}dt^2 + r^2\bigg[ d\varphi
+ \frac{2\rho}{lr^2}dt\bigg]^2 + \frac{l^2d\rho^2}{4\rho^2} \quad
(r^2 = l^2a_+\rho^2 + 2\rho + l^2d)\,. \ee This metric is in ADM
form with the square lapse and the shift given by $N^2 =
4\rho^2/l^2r^2$, $N^{\varphi} = 2\rho/lr^2$.

2) For $m^2l^2 = 7/2$, the solution (\ref{Fr}) with $a_- = 0$,
$a_+>0$, $p_+ = 3/2$, and $d= 0$ leads (again after transforming to
the basis (\ref{ab1})) to the metric \be\lb{n1} ds^2 =
-\frac{4x^2}{l^2(2 +l^2a_+x)}dt^2 + (2x^2 + l^2a_+x^3)\bigg[
d\varphi + \frac2{l( 2+l^2a_+x)}dt\bigg]^2 + l^2\frac{dx^2}{x^2} \ee
($x = \rho^{1/2}$). This spacetime is free from naked CTC.
The point horizon at $x = 0$ hides a timelike causal
singularity at $x = -2/a_+l^2$. At spacelike infinity, the
two-dimensional metric reduced relative to $\partial_{\varphi}$ goes
as \be\lb{n2} ds^2 \simeq \frac4{l^2}y^{-2}(-dt^2 + dy^2)\,, \ee
with $y = (x/l^2a_+)^{-1/2} \to 0$, so that spacelike infinity is
conformally timelike.

3) For the value $m^2l^2 = 1/2$ ($p_+=1$), the ``logarithmic"
solution (\ref{F1}) leads to three black hole subcases:

\noindent a) If $a\neq0$ and $d\neq0$, the metric in the basis
(\ref{ab1}) is similar to (\ref{null}) with \be\lb{r21} r^2 =
al^2\rho\ln|\rho/\rho_0| + 2\rho + dl^2\ln|\rho/\rho_0|\,. \ee
Necessary conditions for the absence of naked CTC are $a>0$ (no CTC
at infinity), and $d<0$ (no CTC
near the horizon). This metric is ``almost" asymptotically $AdS_3$
(the asymptotic behavior overshoots that of $AdS_3$ by a logarithmic
factor).

\noindent b) If $a\neq0$ and $d=0$ (solution (\ref{F12})), the only
difference with the preceding case is that now \be\lb{r212} r^2 =
al^2\rho\ln|\rho/\rho_1| + \ol{d}l^2\,, \ee with $\rho_1 =
\rho_0e^{-2/l^2}$. This is free from naked CTC provided $a>0$ and
$\ol{d}>a\rho_0/e$.

\noindent c) If $a=0$ and $d\neq0$, the ADM shift $N^{\varphi}$ goes
at spatial infinity to a constant in the basis (\ref{ab1}). A metric
with an asymptotically vanishing shift function may be obtained by
transforming back to the frame (\ref{ab}). This metric, \ba\lb{r213} ds^2
&=& -\frac{4\rho^2}{l^2r^2}d\ol{t}^2 + r^2\bigg[ d\ol\varphi -
\frac{dl\ln|\rho/\rho_0|}{r^2}d\ol{t}\bigg]^2 +
\frac{l^2d\rho^2}{4\rho^2} \nn\\ && \qquad (r^2 = 2\rho +
dl^2\ln|\rho/\rho_0|)\ea is free from naked CTC provided $d<0$.
It is asymptotically $AdS_3$ in the weak sense of log gravity
\cite{GJ,HMT,MSS}, and develops a timelike causal singularity at
some negative $\rho_2 > -\rho_0$.

Note that (\ref{Fr}) with $a_- = 0$ reduces to (\ref{Ftmg}), so that
the black holes (\ref{null}) and (\ref{n1}) also solve the equations
of TMG (the fact that these are black holes was overlooked in
\cite{part}), while the black holes (\ref{r212}) and (\ref{r213})
reduce (after appropriate coordinate transformations) to the black
hole solutions of TMG (\ref{mul-1}) and (\ref{chir}) given in
\cite{gaston}.

\setcounter{equation}{0}
\section{Physical parameters}
Now we compute the physical parameters of these black holes. 
By applying Wald's general formula \cite{wald93} the black hole
entropy was found in \cite{newmass} to be given in NMG by \be S
=\frac{\pi}{2G}\left(r - \frac{r}{m^2}\left[(g^{00})^{-1}\R^{00} +
g_{22}\R^{22} - \frac34\R\right]\right)_h \,, \ee evaluated on the
horizon $\rho=0$, with $r^2 = g_{\varphi\varphi}$. Because of the
possible presence of logarithmic factors, we evaluate this more
carefully than in \cite{newmass}. From the expressions of the Ricci
tensor components given there, we find (for $\zeta=1$) \be\lb{entro}
S = \frac{\pi}{2G}\left[\left(1 + \frac1{2m^2}\left[(\X\cdot\X'') -
\frac14(\X'^2)\right]\right)r -
\frac1{2m^2}R^2r^{-1}(r^2)''\right]_h\,. \ee For the ansatz
(\ref{sd}), this leads to \be\lb{S} S =
\frac{\pi}{2G}\left[\left(1-\frac1{2m^2l^2}\right)r_h - \frac2{m^2}
\left(r^{-1}\rho^2F''(\rho)\right)_h\right]\,. \ee For the null
warped $AdS_3$ black holes (\ref{null}), this formula gives the
Bekenstein-Hawking entropy renormalized by a factor $16/17$
\cite{newmass}. In the case of the black holes (\ref{n1}) and
(\ref{r212}), the formula (\ref{S}) yields straightforwardly \be S =
0\,. \ee The case of of the black holes (\ref{r21}) and (\ref{r213})
is more delicate. In this case, $\rho^2F'' = a\rho-d$ does not
vanish on the horizon but goes to a constant, however this is
suppressed by the prefactor $r^{-1}$ which goes to zero as an
inverse logarithm, while the first term in (\ref{S}) does not
contribute because $m^2l^2 = 1/2$ for this case, so that the net
entropy is again zero.

Provided
\be\lb{as}
\mbox{\rm lim}_{\rho\to\infty}\left(\rho^{-1}F(\rho)\right) = 0\,,
\ee
the metric (\ref{sd1}) is asymptotically $AdS$. In that case we can 
for large $\rho$ linearize the metric around the BTZ vacuum as
\be
g_{\mu\nu} = \bar{g}_{\mu\nu} + g^L_{\mu\nu} \,,
\ee
and use the 
Abbott-Deser-Tekin (ADT) approach \cite{AD,DT02} to compute 
the mass and angular momentum of our black holes. The ADT conserved 
charge associated with a background Killing vector $\xi$ has been 
computed for NMG in \cite{liusun2}. Using the coordinate-free parametrization 
(\ref{sd}), such that $\bar{\X} = \b\rho$, $\X^L = \a F(\rho)$, and 
evaluating the covariant derivatives and Ricci tensor components with the 
help of the formulas given in Appendix B of \cite{adtmg}, we 
obtain\footnote{The sign of our $Q(\xi)$ is opposite to that used in 
\cite{liusun2}.} from Eqs. (2.25) and (2.27) of \cite{liusun2} the Killing 
charge
\be\lb{Q} 
Q(\xi) = \frac1{Gm^2l^3}\left[\rho^3F''' + \rho^2F'' +
\frac14\left(\frac12-m^2l^2\right)(\rho F'-F)\right]\left([\a]\xi\right)^0\,, 
\ee 
where $[\a]$ is the matrix 
\be\lb{vecmat} [\a]  \equiv \left(\begin{array}
[c]{cc} -\alpha^Y & -\alpha^T + \alpha^X \\ \alpha^T + \alpha^X & 
\alpha^Y \end{array}\right)\,.
\ee
The charge (\ref{Q}) is a constant of the motion by virtue of (\ref{mast}). 
For the generic solution (\ref{Fr}) or (\ref{Fi}), as well as in the 
special cases (\ref{F0}) and (\ref{F12}), the evaluation of this charge 
leads to 
\be\lb{Qgen} 
Q(\xi) = \frac1{4Gm^2l^3}\left(m^2l^2-\frac12\right)d\,\left([\a]\xi\right)^0
\,, 
\ee 
while for the special solution (\ref{F1}) ($m^2l^2=1/2$), we find 
\be\lb{Q1} Q(\xi) = \frac2{Gl}d\,\left([\a]\xi\right)^0\,. \ee

These values of the charge were derived under the assumption (\ref{as}) 
which is valid, in particular, for the generic solution (\ref{Fr}) with 
$a_+ \neq 0$, provided $m^2l^2 < 1/2$ and for the solution
(\ref{F1}) ($m^2l^2 = 1/2$) with $a=0$. We will here assume that 
they can be extrapolated to the cases $m^2l^2 > 1/2$ with $a_+\neq 0$ and 
$m^2l^2 = 1/2$ with $a\neq0$. A more 
satisfactory derivation would require an extension of the ADT approach to the 
case of massive gravity with non-constant curvature backgrounds, similar to 
that carried out in \cite{adtmg} for topologically massive gravity.
Choosing $\xi$ to be one of two Killing vectors $\xi_{(t)} = (-1,0)$ and
$\xi_{(\varphi)} = (0,1)$, we obtain the mass and angular momentum of the 
various black holes of the previous section\footnote{The super-angular 
momentum approach of \cite{black} as applied to NMG in \cite{newmass} leads
to the same results.}:

 1) In the null warped black hole case (\ref{null}), \be M
= 0\,, \quad J = \frac{4dl}{17 G}\,, \ee in accordance with the
results of \cite{newmass}.

 2) For the black holes (\ref{n1}), \be M = J = 0\,. \ee

 3ab) For the black holes (\ref{r21}) and (\ref{r212}), \be
M = 0\,, \quad J = \frac{2dl}{ G}\,. \ee

 3c) For the black holes (\ref{r213}), \be\lb{MJlog} M =
\frac{2d}{ G}\,, \quad J = \frac{2dl}{ G}\,. \ee Because the absence 
of naked CTC constrains $d<0$, the mass is
negative for a positive Newton constant $G$, and vanishes as it
should in the limit $d \to 0$ of the extreme BTZ black hole for
$m^2l^2 = 1/2$ \cite{newmass}. The values (\ref{MJlog}) may be
compared with the corresponding values for the same black holes as
solutions for TMG. Evaluating for the ansatz (2.12) with (2.10) the
TMG super-angular momentum $\J$ \cite{adtmg}, we obtain 
\be 
\J = \frac{4}{\mu l^2}\left[-
\rho^2F'' + \frac12(1-\mu l)(\rho F'-F)\right]\a\,, 
\ee 
leading for $\mu l = 1$, $F(\rho)$ given by (\ref{chir}) and $\a$ as in
(\ref{ab}) to \be M = \frac{d}{2 G}\,, \quad J = \frac{dl}{2 G}
\ee (the correction $\Delta M$ coming from the last three terms of Eq. (3.15) 
of \cite{adtmg} vanishes in the present case). These values agree with those 
of \cite{gaston} (Eq.(24), where $k$ is our $d$) up to a factor 2/3.

Finally the Hawking temperature and the horizon angular velocity, computed from
the metric in ADM form, are 
\be 
T_{H} =\frac{1}{4\pi}\zeta
r(N^2)'|_h\,, \quad \Omega_{h}= -N^{\varphi}|_h\,. 
\ee 
The resulting Hawking temperatures vanish for all our black holes, $T_H = 0$.
The horizon angular velocities vanish for the black holes
(\ref{null}) with $d>0$ and the black holes (\ref{r21}) and
(\ref{r212}), while $\Omega_h = -2/lk$ for the black holes
(\ref{null}) with $d=0$ and the black holes (\ref{n1}), and
$\Omega_h = l^{-1}$ for the black holes (\ref{r213}). It follows that
the first law of black hole thermodynamics, which in the case of vanishing 
Hawking temperature reduces to 
\be\lb{first0}
dM = \Omega_{h}dJ\,.
\ee 
is satisfied trivially (both sides vanish) for the black holes (\ref{null}),
(\ref{n1}),(\ref{r21}), and (\ref{r212}), and non trivially ($M = \Omega_hJ$)
for the black holes (\ref{r213})  

\section{Discussion}
In this paper, we have investigated solutions of new massive gravity
with two commuting Killing vectors, one of which is null, with a
special emphasis on black hole solutions. In addition to the
well-known extreme BTZ black holes, we found several black hole
types. The first of these includes the black holes (\ref{n1})
($m^2l^2 = 7/2$) and (\ref{r212}) ($m^2l^2 = 1/2$), both with $d=0$
(so that, as shown at the end of Sect. 2, they have a third local
Killing vector). Because all their physical characteristics
(entropy, mass and angular momentum) vanish, these are not genuine
black holes. A second family ($m^2l^2 = 17/2$) includes the null
warped black holes $\beta^2=1$ of \cite{tmgebh}, with metric
(\ref{null}) enjoying a local $sl(2,R)\times R$ isometry algebra.
These have $\partial_t$ as null Killing vector, are massless, but
have a nonzero angular momentum. The third family ($m^2l^2 = 1/2$),
with metric (\ref{r21}), has similar properties, but only two local
isometries.

The most interesting fourth black hole type (also $m^2l^2 = 1/2$,
corresponding to the value of the bare cosmological constant $\L =
-3m^2$ from (\ref{ell})) differs from the preceding by the fact that
it is asymptotically $AdS_3$ in the sense of log gravity
\cite{GJ,HMT,MSS}. This implies that in the basis (\ref{ab1})
appropriate to the other black hole types, the ADM shift function
$N^{\varphi}$ does not vanish at infinity. After transforming to the
basis (\ref{ab}) in which $N^{\varphi}(\infty)=0$ (which transforms
the null Killing vector to $\partial_v$), we obtained a continuum of
black hole states (\ref{r213}) with mass $M$ and angular momentum
$J=lM$, above the massless extreme BTZ family as ground state. These
properties are similar to those of the ``log" solutions of TMG at
the chiral point $\mu l= 1$ found in \cite{gaston}. Further work is
needed to understand the implications of these solutions on the
consistency of new massive gravity at the critical value $m^2l^2 =
1/2$.

\renewcommand{\theequation}{A.\arabic{equation}}
\setcounter{equation}{0}
\section*{Appendix}

We discuss here the geodesic equation (\ref{geo}) in the case of the
generic solution (\ref{Fr}). Assuming that $a_+$ and $a_-$ do not
both vanish, we must distinguish between four possibilities for the
leading near-horizon behavior of the effective potential:
\begin{enumerate}

\item $a_- \neq 0$ and either $p_- < 0$, or $p_- > 0$ and $d = 0$. The leading term in
$F(\rho)$ is $a_-\rho^{p_-}$, with $a_- > 0$ (if $a_- < 0$ the
spacetime is geodesically complete, with naked CTC \cite{part}), so
that the affine parameter is proportional to $x = \rho^{(2-p_-)/2}$
near the horizon. The geodesic equation (\ref{geo}) can be rewritten
in terms of the adapted radial coordinate $x$ as \be \lb{geo1}
\dot{x}^2 - l^2P_+^2x^{-2p_-/(2-p_-)}F(x^{2/(2-p_-)}) -
2P_+P_-x^{2-2/(2-p_-)} + \frac{4\varepsilon}{l^2}x^2 = 0\,. \ee The
geodesics can be extended across the horizon $x = 0$ if the
effective potential in (\ref{geo1}) is analytical in $x$
\cite{sig,cold}. A necessary condition for this is that the exponent
$2-2/(2-p_-)$ be integer. However $p_-<1/2$ leads to
$2/(2-p_-)<4/3$, so that this exponent can be integer only for
$p_-=0$, corresponding to the special case (\ref{F1}) or (\ref{F12})
(see below). The conclusion is that generically geodesics terminate
at the singular horizon $x=0$.

\item $d \neq 0$ and either $a_- = 0$, or $a_- \neq 0$ and $p_- > 0$. The leading
term in $F(\rho)$ is constant, so that the adapted radial coordinate is
$x = \rho$, and the horizon is regular provided the function $F(\rho)$ in
(\ref{Fr}) is analytic. For $a_- \neq 0$, this is not possible because
$p_-<1/2$ is not integer. For $a_- = 0$, $p_+$ must be integer, $p_+ = n$.
The case $n=1$ corresponds to $p_- = 0$ (see below). In the case $n=2$ ($m^2l^2
= 17/2$), (\ref{sd}) reduces to the quadratic ansatz $\X = \tilde{\a}\rho^2 +
\tilde{\b}\rho + \tilde{\c}$, with $\tilde{\c} \propto \tilde{\a}$, leading
(after an appropriate coordinate transformation) to null warped black
holes (Eq. (3.17) of \cite{tmgebh} with $\beta^2 = 1$, $\rho_0 = 0$,
$c=l^2a_+$,  $\omega = 1$ and $u = dl^2/2$ in (3.20)). Finally,
in the case $n > 2$ and $a_+ > 0$ ($a_+ < 0$ leads to naked CTC), the
affine parameter is for large $\rho$ proportional to $\rho^{(2-n)/2}$, so that
geodesics terminate at infinity.

\item $a_- = d = 0$, $a_+ \neq 0$ and $p_+ < 1$. The leading term in
$F(\rho)$ is $a_+\rho^{p_+}$, with $a_+ > 0$ ($a_+ < 0$ leads again
to a geodesically complete spacetime with naked CTC) so that the
affine parameter is proportional to $x = \rho^{(2-p_+)/2}$ near the
horizon. The geodesic equation rewritten in terms of $x$ is of the
form (\ref{geo1}) with $p_-$ replaced by $p_+$. From $1/2 < p_+ <
1$, we find  $4/3 < 2/(2-p_+) < 2$, so that the exponent
$2-2/(2-p_+)$ is not integer, and the horizon is singular.

\item $a_- = d = 0$, $a_+ \neq 0$ and $p_+ > 1$. Now the leading term in
the effective potential is linear in $\rho$, so that the affine
parameter is proportional to $x = \rho^{1/2}$ near the horizon. The
geodesic equation rewritten in terms of $x$ is \be \lb{geo2}
\dot{x}^2 - l^2P_+^2a_+x^{2(p_+-1)} - 2P_+P_- +
\frac{4\varepsilon}{l^2}x^2 = 0\,. \ee The effective potential is
analytical provided $p_+ = n/2 +1$, with $n$ a positive integer. For
$n=1$ ($m^2l^2 = 7/2$, corresponding to $\L/m^2 = - 15/49$), we
obtain (after transforming to a rotating frame) the metric
(\ref{n1}). For $n=2$ ($p_+ = 2$), the resulting metric is (again
after transforming to a rotating frame)a special case of the null
warped black holes $\beta^2 = 1$ of \cite{tmgebh} (with $u=0$ in eq.
(3.20) of \cite{tmgebh}). For $n > 2$, the affine parameter is for
large $\rho$ proportional to $\rho^{(2-n)/4}$, so that again
geodesics terminate at infinity.

\end{enumerate}

There remains the case of the special solutions (\ref{F0}) and
(\ref{F1}) or (\ref{F12}). The solution (\ref{F0}) corresponds to
the degenerate case $p_+=p_-=1/2$, which from the previous analysis
cannot possibly lead to regular black holes. In the case of the
solution (\ref{F1}), the affine parameter is for $d<0$ proportional
to $x = \rho(-\ln|\rho/\rho_0|)^{-1/2}$ near the horizon. The
geodesic equation rewritten in terms of $x$ will contain only
integer powers of $x$ and powers of $y=-\ln|\rho/\rho_0|$, which is
positive on both sides of the horizon, so that geodesics can be
continued across the horizon. The same conclusion holds for the
solution (\ref{F12}) with $\ol{d}>0$ and $x=\rho$.

\end{document}